\documentclass[aps,pra,11pt,reprint,longbibliography]{revtex4-1}
\usepackage{lineno}
\usepackage{amssymb}
\usepackage{amsmath}
\usepackage{bigints}
\usepackage{natbib}
\usepackage{ifpdf}
\usepackage{graphicx}
\usepackage{hyperref}
\hypersetup{pdfauthor={Sebastien Berujon and Eric Ziegler},pdftitle={Near-field speckle-scanning-based x-ray tomography}}

\DeclareMathOperator*{\argmax}{arg\,max}

\begin{document}

\title{Near-field speckle-scanning-based x-ray tomography}
\author{Sebastien Berujon} \email[]{berujon@esrf.eu}
\affiliation{European Synchrotron Radiation Facility, BP 220, F-38043 Grenoble Cedex 9, France}
\author{Eric Ziegler}
\affiliation{European Synchrotron Radiation Facility, BP 220, F-38043 Grenoble Cedex 9, France}

\date{\today}

\begin{abstract}
We previously demonstrated near-field speckle scanning based x-ray imaging to be an easy-to-implement phase sensing method capable of providing both high sensitivity and high resolution. Yet, this performance combination could only be achieved at the cost of a significant number of sample exposures and of extensive data acquisition time, thus tempering its implementation for tomography applications. Here, we show ways of drastically lowering the number of exposures for the speckle scanning method to become attractive for computed tomography imaging. As the method presented can cope with a high divergence beam, it is also expected to attract the attention of the laboratory sources community.
\end{abstract}

\maketitle

\section{Introduction}

The famous image of R\"ontgen's wife's hand, where the ring and bones could easily be identified, gave rise to a century of ever more impressive x-ray imaging deeds. X-ray absorption contrast imaging has become an essential tool for the development of human knowledge in area such as medical applications, material science, historical heritage and even security screening. Further progress now requires a greater level of sophistication, with x-ray phase contrast imaging being one of these advances that foster a boost in the x-ray imaging domain. Within the last two decades, supported by the advent of coherent x-ray sources and the wish to image objects with ever higher contrast, sensitivity together with a better resolution, scientists devoted some of their efforts to map the refractive part $\delta$ of the optical index $n = 1-\delta -i\beta$ in samples \cite{davis1995,momose1996,nugent1996,cloetens1996,snigerev1996}. Being orders of magnitude larger than its counterpart absorption factor $\beta$, the contrast induced by the $\delta$ factor permits to obtain information that was unreachable before.

Within a couple of decades a few approaches arose from the early demonstration of phase contrast imaging in the late 1990s, concomitant with the advent of modern synchrotrons and laboratory sources. Roughly speaking, the available methods can be split into two categories with, on one hand, the ones sensitive to the Laplacian of the x-ray beam phase \cite{nugent1996,paganin2002,cloetens1999} and, on the other hand, the ones sensitive to the beam phase gradient \cite{david2002,pfeiffer2007,olivo2007,brun2012}. The techniques of this second category benefit from a high sensitivity to slowly varying electronic density and from a moderate need of \emph{a priori} assumptions on the sample composition. They are also less demanding in term of source size, making them more adaptable to laboratory sources.

Among the repertoire of x-ray phase gradient sensitive methods, the interest for near-field speckle based ones is growing very rapidly due to their ease of implementation and suitability to sources presenting moderate transverse coherence \cite{cerbino2008,berujon2012prl,morgan2012}. These methods rely on the modulation of the x-ray light with a random object, either with an absorption mask or by the near-field speckle effect when using a coherent beam and weakly absorbing objects \cite{cerbino2008}. Due to the nature of this random object or diffusor, e.g. a piece of sandpaper or a filtering membrane, such a wavefront modulator is available at a negligible cost. Besides, the identification of various processing schemes provided options to best tune the method in term either of sensitivity, resolution or experimental requirements \cite{berujon2013apl,berujon2016prap}.

The near-field speckle-based processing methods differ by their numerical processing and by the mode of data acquisition they require to recover the phase from a sample image. With simple setups, the x-ray speckle tracking (XST) \cite{berujon2012prl,morgan2012} and speckle vector tracking (XSVT) \cite{berujon2015pra} approaches demonstrated the possibility of three-dimensional (3D) numerical volume reconstruction \cite{wang2015,berujon2016prap} within a computed tomography (CT) process. Nonetheless, it is the x-ray speckle scanning (XSS) method that offers the best performance combination in terms of sensitivity and resolution. This method was shown to be a generalization of the grating interferometer used in the phase stepping mode \cite{berujon2012pra}. Whilst the angular sensitivity of XSS can approach the single nanoradian, the spatial resolution can be pushed down to a scale smaller than the imaging detector pixel size. This is possible when using setups with x-ray magnifying optics or with source presenting large divergence. Such an aspect makes the method greatly valuable at laboratories where large pixel detectors are used in combination with compact x-ray sources possessing a large solid angle emission, the setups often generating a large magnification ratio.
Yet, the main current drawback of the XSS method lies in the large number of exposures to the x-ray beam necessary for each projection to recover the phase gradient induced by the sample presence. Using a two-dimensional (2D) approach, more than 100 images are usually required in each projection scan while a one dimensional (1D) approach providing a high sensitivity in only one direction requires about 40 exposures per projection \cite{wang2016scr,wang2016apl}.

In this article, we demonstrate the recovery of the full 2D x-ray differential phase gradient induced by a sample with high accuracy with as little as 25 sample exposures. In a CT process, this number is even reduced further by using suitable interlaced schemes, as we experimentally demonstrate. Such levels of sophistication make the data acquisition time readily CT-compatible. In a first theoretical part we explain the approach and its numerical implementation and provide ways of simplifying the data acquisition and processing. Then, in the experimental part, applications of the methods on real samples are presented, including both radio-projection imaging and 3D CT.

\section{Theory}
\subsection{Basis}
The fundamental principle of the speckle based techniques is very much the same as the one employed to understand a shearing interferometer \cite{pfeiffer2007} or a coded aperture system \cite{olivo2007}. In all of these techniques, the wavefront is modulated in intensity with a phase or a transmission object located upstream or downstream of a sample. As a matter of fact, the method explained further can also be seen as a coded aperture system in which the coding object is a random mask. This was experimentally applied for instance in \cite{wang2016high} where absorption contrast instead of interference contrast was used.

A random wavefront modulating mask can be made of small absorbing features or of the interference generated by grains placed in a coherent beam, known as speckle. The \emph{sine qua non} condition for x-ray speckle visibility is the use of a coherent light source such as a synchrotron or a microfocus source. The main asset of the speckle modulator lies in the very efficient modulation it can produce with little or no photon absorption. Moreover, in the hard x-ray regime, the extent of the speckle near-field region is much greater than with longer wavelength light, typically up to several meters. This means that the speckle pattern distortion upon propagation will be ruled only by the wavefront deformation over such distances \cite{cerbino2008}. In parallel, simulations showed that usable x-ray near-field speckle could be obtained with spectral bandwidths as large as 20\% \cite{zdora2015}.

Let us consider a wavefront modulator, here a diffusor or a membrane, movable upon time and located at the position $\tau(t)$ and we note $I_S(P,\tau)$ and $I_R(P,\tau)$ the intensities collected on the detector at the pixel position $P = x\mathbf{e_x} + y\mathbf{e_y}$, respectively in the presence of a sample (sample scan) into the beam and in absence of it (reference scan).

In projection imaging, the transmission and phase shift induced by the sample on the photon beam are respectively,
\begin{equation}
T(P) = \exp\left(-2k\int \beta(P,z)dz\right)
\end{equation}
and phase
\begin{equation}
\phi = -k\int \delta (P,z) dz
\end{equation}
with $k = 2\pi /\lambda$ the wavenumber, and $\lambda$ the wavelength. Whilst $T$ is the signal easily accessed with an imaging detector and used in absorption contrast based x-ray imaging, the access to the quantity $\phi$ is more subtle.

Using the derivation of Munro \cite{munro2013} for a coded aperture system, or using simplifications of the Transport of Intensity Equation, we have the relationship:
\begin{equation}
I_S(P,\tau)\approx T(P) I_R\left(P+ \frac{d}{k} \nabla\phi/M,\tau\right)
\label{eq:princ}
\end{equation}
where $d$ is the distance from the sample to the detector, and $M$ the setup magnification. This equation holds true for a monochromatic beam or for a system that can be defined by an equivalent energy for a beam with a broader spectrum.

The methods exposed in the literature concerning x-ray grating interferometry operating in phase shifting mode, some coded aperture approaches as well as the speckle scanning approach differ from each other in the way they invert this last relation for extracting $T$ and $\phi$ from data collected whilst scanning the modulator.

Already, from Eq.~\ref{eq:princ}, one can see that upon pixel wise normalization of $I_S$ and $I_R$ to $\widehat{I_S}$ and $\widehat{I_R}$, and noting $\nabla$ the del operator, we have:
\begin{equation}
\begin{aligned}
\widehat{I_{S}}(P,\tau) & \approx \widehat{I_{R}}\left(P+\frac{d}{k} \nabla\phi/M,\tau\right)
\label{eq:tau}
\end{aligned}
\end{equation}
which is equivalent to
\begin{equation}
\begin{aligned}
\widehat{I_S}(P,\tau) & \approx \widehat{I_R}\left(P,\tau- \frac{d}{k} \nabla \phi/M\right)\\& \approx\widehat{I_R}\left(P,\tau-\Delta \tau\right)
\end{aligned}
\end{equation}
if we choose to move $\tau$ at a constant speed $\eta$ \cite{berujon2014}.
Thus, the principle common to all scanning phase gradient sensitive techniques is to recover $\Delta \tau$ to then be able to derive the quantity $\nabla\phi$.

\subsection{Sparse sampling for 2D XSS\label{sec:sparse}}

Here we present applications of the XSS technique based on 2D scans. In previous papers devoted to the description and explanation of the method, dozens of images were recorded for the phase recovery of a single projection. Here, a much sparser sampling during data acquisition with the sample inserted in the beam allows a valuable optimization of the exposure time to x rays. This reduction on the number of acquisitions necessary for correct high resolution phase recovery is possible through a wise selection of the diffusor scanning positions.

\begin{figure}
\ifpdf
\includegraphics{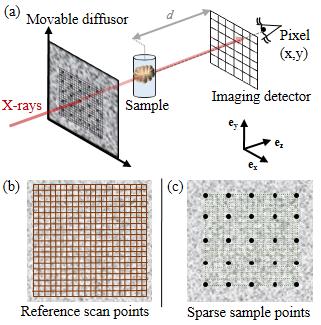}
\else
\includegraphics{scanning2.eps}
\fi
\caption{(Color online) (a) Sketch of the XSS setup: the x-ray beam passes through the movable diffusor and the sample before impinging on the imaging detector. (b) Mesh grid of the membrane positions for reference scan recording. (c) Mesh grid of the membrane positions, black points, where data are recorded during the sample scans. The green grid is represents the bilinearly interpolated data points. Here $(\varrho_1,\varrho_2)=(4,4)$\label{fig:concept}}
\end{figure}

Figure~\ref{fig:concept}.(a) sketches the setup of the XSS method. Therein, a sample is placed into the x-ray beam at a close distance from a movable diffusor, generally a membrane. The latter, mounted either upstream or downstream of the sample, generates small random intensity features recorded by the detector. The diffusor can be scanned transversally to the x-ray beam propagation direction. During the XSS data acquisition, the diffusor is translated according to a mesh grid of positions with a regular micrometer step noted $\eta$ while an image is acquired at each position $\tau = X\mathbf{e_x} + Y\mathbf{e_y}$. Next, a reference data set is generated by repeating the same acquisition scan, this time with the sample away from the beam. As a result, for each pixel $P$, two data arrays $I_{S}(P,\tau)$ and $I_{R}(P,\tau)$ are available.

The rigid shift in position $\mathbf{\Delta \tau} $ between these two similar patterns acquired in $P$ can be recovered after normalization as: 
\begin{equation}
\mathbf{\Delta \tau} = \argmax_{v} \int \widehat{ I_S}(P,\tau) \widehat{ I_R}(P,\tau + v)d\tau
\label{eq:corr}
\end{equation}
which can be tracked with a substep accuracy \cite{pan2009}.
This vector is directly proportional to the local wavefront gradient $\nabla W_P$ and differential phase $\nabla \phi_P = k\nabla W_P$ through the relationship induced by Eq.~\ref{eq:tau}:
\begin{equation}
\nabla \phi_P = k\nabla W_P \approx M\frac{k}{d}\mathbf{\Delta \tau}
\end{equation}
Noting $R$ the distance from the source to the diffusor or the sample, whichever is the closest and, $l$ the distance from the diffusor to the sample, we have \cite{berujon2015pra}:
\begin{equation}
M = (R+l+d)/R
\end{equation}

In previous works, both patterns $I_{S}(P,\tau)$ and $I_{R}(P,\tau)$ were built following the most basic sampling:
\begin{equation}
(X,Y) = \sum_{[\![-q:q]\!]}( \delta_x-\eta q, \delta_y-\eta q)
\end{equation}
where $q\in \mathbb{N}$ and $\delta_{x,y}$ denote in this case the Dirac distribution. Such sampling makes the requirement on the number of images relatively high and more importantly unnecessary since we aim at recovering an injective function through Eq.~\ref{eq:corr}.
Hence, to reduce the number of sample exposures, we suggest building the correlated signal $I_{S}(P,\tau)$ from a much sparser sampling scheme than for $I_{R}(P,\tau)$. Figures~\ref{fig:concept} (b) and (c) show schematically the dissimilar sampling of the two signals. Whilst $I_{R}$ is built from the images collected as usual over a fine mesh scan of the diffusor, $I_{S}$ is built from fewer points along a mesh with the following sparser grid:
\begin{equation}
(X,Y)_S = \sum_{[\![-q:q]\!]} \left(\delta_x-\varrho_1\eta q, \delta_y-\varrho_2\eta q\right)
\end{equation}
For further processing, the pixel signal collected for $I_{S}$ is then interpolated by the factors $(\varrho_1,\varrho_2)$ using a bilinear method to generate a virtual mesh sampling grid that matches the one recorded for $I_{R}$. At this stage Eq.~\ref{eq:corr} can be applied for $\phi_P$ to be recovered.

In fact, the sparsity of the sampling performed during the acquisition of $I_{S}$ permits the incorporation of sufficient independent statistics into the cross-correlation process to ensure the accurate recovery of the displacement vectors. Meanwhile, the sampling of outer points in the reference scans warrants the existence of a correlation peak by the recording of patterns with a matching portion in the two scans.

\subsection{Wavefront Laplacian\label{sec:secderiv}}

For a more reliable wavefront recovery, the XSS method provides the possibility of accessing the Laplacian of the wavefront. Although it requires a greater number of sample exposures than for the method described in Sec. \ref{sec:sparse}, this processing mode is of interest when imaging samples generating very turbid wavefronts.
As the setup still involves a sample and a scattering membrane located in front of the imaging detector with a pixel size $p_{ix}$, we keep the notation $I_{S}(P,\tau)$ and $I_{R}(P,\tau)$ as before.

Equation~\ref{eq:princ} permits the recovery of $\nabla \phi$ with a good accuracy especially when the wavefront distortion is smooth and continuous. For strongly varying wavefronts, artefacts may occur upon the 2D integration of $\nabla \phi$. Henceforth, the inclusion into the integration step of the wavefront second derivative through a Taylor expansion helps the algorithm to deal with phase jumps.

For a pair of neighboring pixels $P_1$ and $P_2$ with an interdistance $q p_{ix}, q \in \mathbb{N}$ , we note $\tau_S$ and $\tau_R$ the delay for which the pixel $P_2$ registers the same signal as the one measured in $P_1$ when scanning the diffusor. The absolute wavefront curvatures of the x-ray beam for each state, i.e. with and without the sample inserted into the beam, can be recovered through the relations \cite{berujon2012pra}:
\begin{equation}
\begin{aligned}
\nabla^2 W_S = \frac{1}{d}\left( \frac{\tau_S}{qp_{ix}}-1\right) ,
\nabla^2 W_R = \frac{1}{d}\left( \frac{\tau_R}{qp_{ix}}-1\right)
\end{aligned}
\end{equation}
By linearization in the small angle approximation, the differential second wavefront derivative becomes:
\begin{equation}
\begin{aligned}
\nabla^2 W &= \nabla^2 W_S - \nabla^2 W_R = \frac{1}{dqp_{ix}}(\tau_S - \tau_R)
\end{aligned}
\end{equation}
Using the Taylor expansion:
\begin{equation}
W(P_2) \approx W(P_1) + q p_{ix}\nabla W + \frac{q^2 p_{ix}^2}{2} \nabla^2 W
\end{equation} we can operate the wavefront integration using a matrix inversion \cite{berujon2015josr}, i.e. we use:
\begin{equation}
\phi \approx M\frac{k}{d} \bigintsss \left( \mathbf{\Delta \tau}+\frac{1}{2}(\tau_S - \tau_R)\right)d\mathbf{P}
\end{equation}
Such processing implies that $I_{S}(P_1,\tau)$ and $I_{S}(P_2,\tau)$ contain partially similar signals, putting hence requirements on the scanning range. During the data collection, the scan length must be long enough; at least many times $qp_{ix}/M$ .

\subsection{ Vector tracking generalization\label{sec:gen}}

The XSS method of the previous sections could be further reduced, both conceptually and numerically, to lighten up the computational processing routine. Whilst an interpolation step was used in Sec.~\ref{sec:sparse}, we show here an approach based on vectors, as in the case of the XSVT method \cite{berujon2015pra}.

\begin{figure}
\ifpdf
\includegraphics{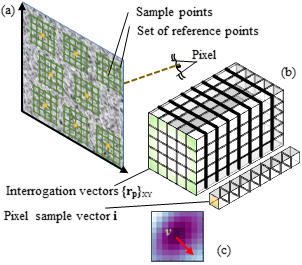}
\else
\includegraphics{vectors.eps}
\fi
\caption{(Color online) Conceptual sketch of the data collected in one single pixel during the reference and sample scans. (a) The green and yellow dots are markers of the membrane position seen by the pixel during the reference and the sample scans, respectively. (b) Vectors built for one pixel using the data collected during the multiple reference scans and stacked together. (c) Correlation peak calculated using the vector built with the sample data and correlated with the reference vectors.\label{fig:gensch}}
\end{figure}

As sketched in Fig.~\ref{fig:gensch}, a set of $\left\{\tau_i\right\},{i\in[\![1:N]\!]}$ membrane positions randomly distributed and called "sample points" are defined; these positions, marked with yellow dots in the figure, correspond to the points where data are collected for the scan with the sample present in the beam. For pixel $P$, we build a vector $\mathbf{i}_p$ from the intensities $I_S(P,\tau_i)$:

\begin{equation}
\mathbf{i}_p = \left( I_S(P,\tau_0),...,I_S(P,\tau_i),...,I_S(P,\tau_N) \right)
\end{equation}

Next, the idea is again to solve Eq.~\ref{eq:tau} by finding $\Delta \tau$ such that $\hat{I_R}(P,\tau_i+\Delta \tau) = \hat{I_S}(P,\tau_i)$. For this, we built apart a set of reference vectors by scanning the membrane with mesh scans using a fine regular step $\eta$ around each diffusor position previously defined as sample positions. Thus, by collecting 2D arrays of intensity values for each pixel, we can reorganize these data using the specific order of the sample points and construct a set of reference vectors $\{\mathbf{r}_p\}_{XY}$:

\begin{equation}
\begin{aligned}
\mathbf{r}_p (X,Y)_q &=(  I_R(P,\tau_0 + \Delta \tau_{q}),...,  \\
 & I_R(P,\tau_i + \Delta \tau_{q}),...,I_R(P,\tau_N + \Delta \tau_{q}  ) )
\end{aligned}
\end{equation}

In discrete space, they correspond to the locations: $\Delta \tau_{q}= (X,Y)_q=( \delta-\eta q_1 , \delta-\eta q_2), q=(q_1,q_2)\in [\![-N:N]\!]^2$, which are the coordinates of the fine mesh grid points in the referential of a sample points.
As for XSVT, we make use of the Pearson correlation coefficient $\rho$ \cite{berujon2015pra} to track with a sub-step accuracy the vector $\mathbf{i}_p$ across the reference stack of reference vectors:
\begin{equation}
\mathbf{\Delta \tau} = -\argmax_{(X,Y)}\rho \left[ \mathbf{i_p},\mathbf{r_p}(X,Y) \right]\label{eq:disp}
\end{equation}

The location $\mathbf{v_P}$ of the reference vector providing maximum correlation with the sample vector eventually provides the phase gradient since $\nabla \phi = M\frac{k}{d}\mathbf{\Delta \tau}$.

\section{Experiments}
\subsection{Setup}

\begin{figure}[h]
\ifpdf
\includegraphics{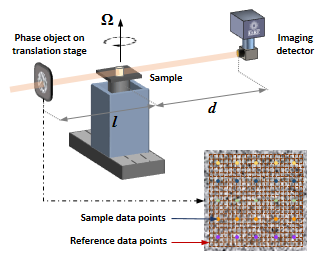}
\else
\includegraphics{setup.eps}
\fi
\caption{(Color online) Top: Experimental setup. Bottom: diffusor transverse position during the data acquisition. In red is the mesh grid defining the membrane positions during the reference scan. The larger colored dots mark the membrane positions when the sample is in the beam: each color represent a set of points corresponding to different projection angles.\label{fig:setup}}
\end{figure}

The experimental setup we used, shown in Fig.~\ref{fig:setup}, is mounted in the second experimental hutch of the beamline BM05 of the ESRF \cite{ziegler2004}. It involves a tomography stage located at $R$ = 55 m from the bending magnet source, that generates x-ray photons by synchrotron radiation from a 0.85 T field magnet on the 6.02 GeV electrons of the storage ring.
Often, the parameters for speckle based imaging must follow some recommendations if one does not want to encounter limitations \cite{aloisio2015,kashyap2016}. For the experiments, pieces of sandpaper with a P220 grit (68 $\mu$m average particle diameter) were stacked and mounted on a piezoelectric translation stage placed at a distance l = 440 mm upstream the sample position. The piezo actuators allow the displacement of the scattering object with a nanometer accuracy over a range of 100 $\mu$m.

Two configurations were tested for the following experiments: one using a monochromatic beam, the other one using a filtered white beam. To generate a monochromatic beam, the beamline double-crystal Si(111) monochromator was used to select photons with an energy of 17 keV and a bandwidth selectivity of $\sim10^{-4}$. For this configuration, the detector was a FReLoN camera coupled to a scintillator and a magnifying optics rendering an effective pixel size of 5.8 $\mu$m. In this case the distance from the sample to the detector was $d=1$ m.

The second operating mode consisted of the continuous spectrum of the BM05 bending magnet source filtered by transmission through the following combination of foils: 0.3 mm of tungsten and 3 mm of aluminum. The detector was a PCO Edge 4.2 camera with an optics rendering a final pixel size of 4.7 $\mu$m. The scintillator was a 200 $\mu$ thick Lu:AG crystal grown by chemical vapor deposition. Such filtering of the bending magnet source spectrum combined with the luminescence efficiency of the scintillator provided a detected beam spectrum centered around 65 keV with a 25 \% bandwidth. At this energy 20 sheets of sandpaper had to be stacked up together for the interference contrast to be observable. The distance from the sample to the detector was $d=2$ m.

The speckle could be generated thanks to the beam transverse coherence of the beam at the diffusor position which is of $\sim10~\mu$m horizontally by $\sim30~\mu$m vertically for E =17 keV. In the filtered beam configuration these values are much decreased due to the smaller light wavelength and the bulk of filters traversed.

\subsection{Data collection and processing}

\begin{figure}
\ifpdf
\includegraphics{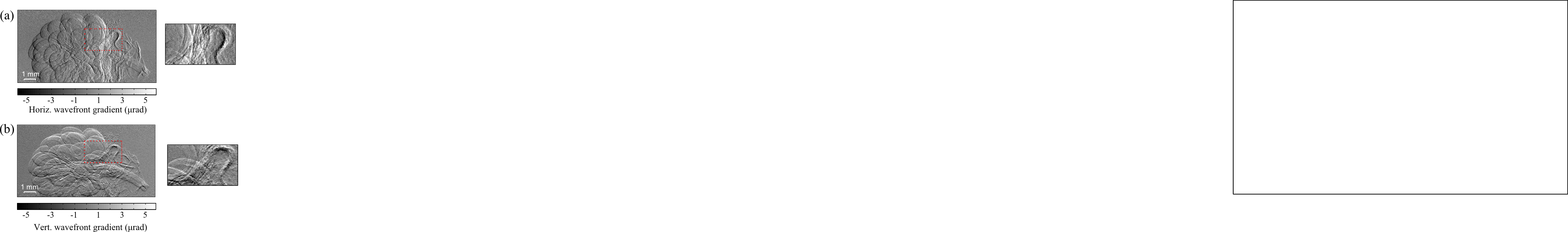}
\else
\includegraphics{muregrad2.eps}
\fi
\caption{(a) Horizontal and (b) vertical wavefront gradients of a blackberry sample. The inset images on the right side are zoomed images of the area marked in red on the larger images.\label{fig:projmure}}
\end{figure}
The samples displayed here consist of a blackberry and a dry vanilla bean. While the blackberry is interesting for the light elements it is made of with varying densities, the vanilla beam is also composed of organic material and contains many high frequency patterns.
\begin{figure*}[ht]
\ifpdf
\includegraphics{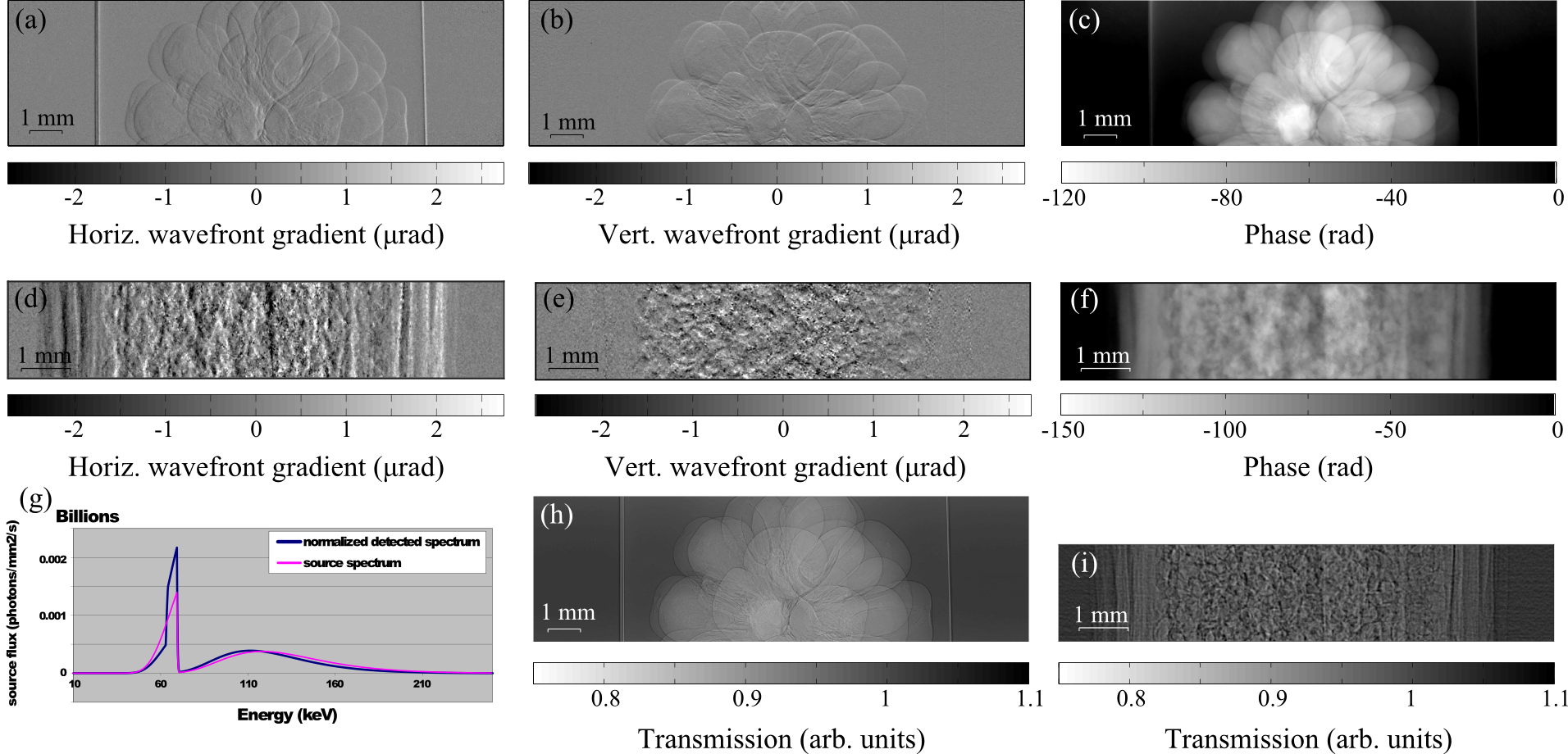}
\else
\includegraphics{Vanilla63keV.eps}
\fi
\caption{(Color online) Images obtained using the broad spectrum beam obtained by filtering of the bending magnet source radiation. (a) Horizontal and (b) vertical wavefront gradients of the blackberry sample. (c) Phase obtained by integration of (a) and (b). (d) Horizontal and (e) vertical wavefront gradients of the dry vanilla beam sample. (f) Phase obtained by integration of (a) and (b). (g,h) Absorption contrast of these same samples. (i) Calculated spectrum of the beam used.\label{fig:pinkbeam}}
\end{figure*}
Before any phase recovery processing the images undergo a normalization operation to account for the varying detector response function and the intensity beam variation. The normalized images $I_N$ are calculated from the recorded images $I_R$ by $I_N = (I_R - <I_R>)/\sigma (I_R)$ where $<.>$ and $\sigma$ denote respectively the mean and standard deviations of the image pixel intensities taken over the full image considered. Such process greatly improves the robustness and accuracy of the phase calculation. It should be pointed out that neither flat nor dark corrections were further applied. Spikes in the images were removed using morphological processing.
The calculation was implemented and tested under MATLAB$^\circledR$. Then, this code was compiled using the software compiler to run the routine in parallel for the CT.

\subsection{Projection imaging\label{sec:proj}}

Projection imaging is the first step before tomography reconstruction. The results of these calculations are shown below.

The wavefront gradient images shown in Fig.~\ref{fig:projmure} for the blackberry sample were obtained with the interpolation based methods of Sec.~\ref{sec:sparse}. A mesh scan of 5$\times5$ images with $\eta = 3~\mu$m was realized to collect the data when the sample was in the beam. For the reference data a mesh scan with a step $\varrho=5$ times smaller was performed. The inset on the right shows the high quality of the data in terms of both sensitivity and resolution. With the XSS mode, the angular sensitivity $\delta \alpha$ can be approximated with $\delta \alpha = M \Delta \tau/d$. While nanoradian sensitivity could be achieved through a high magnification ratio and a large propagation distance, the collimated beam we used in our experiments did not allow such high sensitivity.

For comparison purposes, the phase gradients of this projection were recovered using the previous algorithm with a large number of sample exposures as well as with the two schemes described above, i.e. the one with sparse regular sampling and the other using the random membrane position with matching reference vectors. In an area with no sample, the standard deviation of the phase gradient obtained with 25 exposures and the interpolated scheme of Sec.~\ref{sec:sparse} was of 0.81 $\mu$rad and of 0.91 $\mu$rad when using the scheme of Sec.~\ref{sec:gen}. With 36 sample exposures, these figures fell below 0.30 $\mu$rad, a value that is very near the one achieved with the scheme based on many sample exposures presented in Ref.~\cite{berujon2012pra} with 25$\times$ 25 images and a stability of 0.24 $\mu$rad. Hence, the sparse sampling treatment described above proves to not generate any drastic loss in sensitivity and to provide 2D phase gradient with less sample exposures than it is required to perform 1D XSS.

The same blackberry and a vanilla bean samples were then imaged at higher energy using the bending magnet filtered beam configuration. Figure~\ref{fig:pinkbeam} shows the images calculated with 36 sample exposures and with the generic method of Sec.~\ref{sec:gen}. For both samples and without any \emph{a priori} assumption on the object material, the method was able to accurately retrieve the phase gradient generated by the presence of the sample in the beam. Figure~ \ref{fig:pinkbeam}.(g) shows the detected spectrum, the product of the source spectrum by the scintillator and the filters inserted along the beam. While such phase gradient maps were previously obtained using absorption mask \cite{wang2016high}, the results presented here prove that speckle-based methods are still usable at medium or high energy.

\subsection{Tomography reconstruction}

\begin{figure}[t]
\ifpdf
\includegraphics{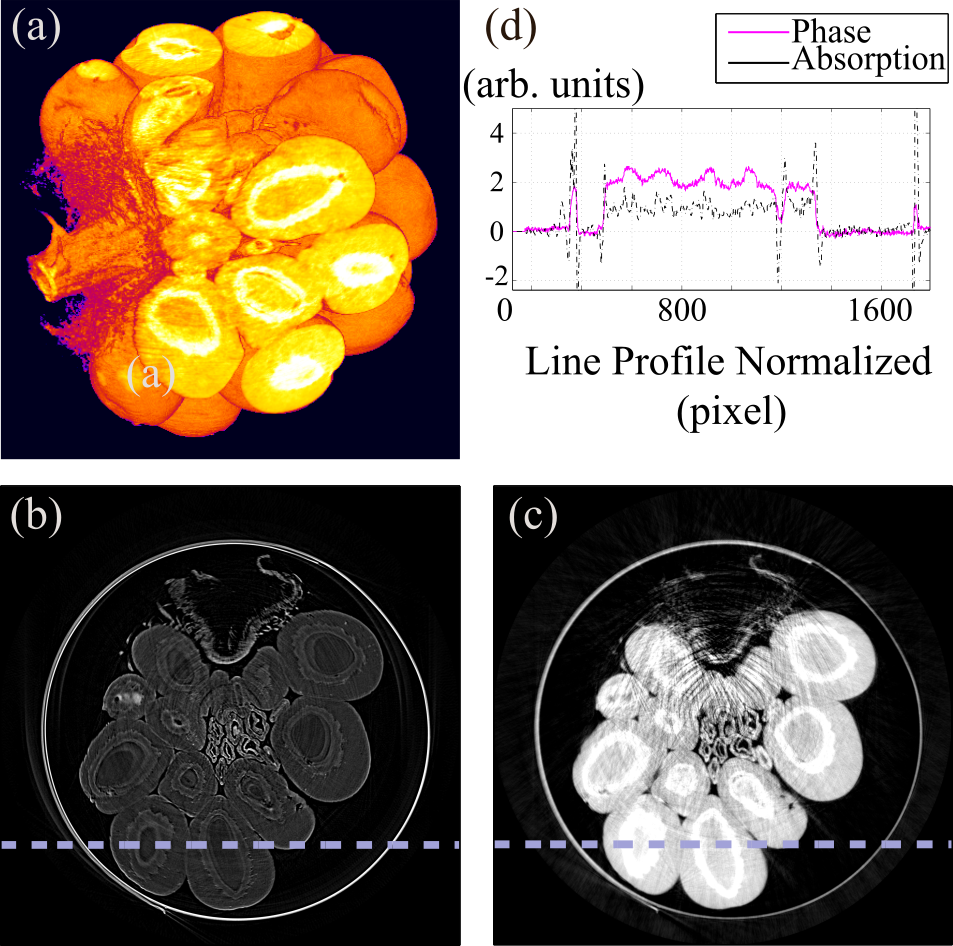}
\else
\includegraphics{mureslice.eps}
\fi
\caption{(Color online) (a) Volume rendering afterb lackberry reconstruction. (b) Absorption contrast and (c) phase contrast slice reconstruction. (d) Cut along the dashed line marked in (b) and (c). \label{fig:mureslice}}
\end{figure}

The samples were then reconstructed using a CT process for volume visualization. Although around 25 images are necessary for accurate recovery of the full 2D wavefront gradient for each projection (cf. Sec.~\ref{sec:proj}), the data acquisition could be hastened by applying an interlaced scheme equivalent to the treatment presented in Ref.~ \cite{berujon2016prap}. It consists of using images of neighboring projections to calculate the one at the angle under consideration. With $\Omega_N$ being the number of angular projection, the top five sample points marked in yellow in Fig.~ \ref{fig:setup} were collected only every five projection $\mod(\Omega_N,5) =0 $. The following five blue points placed underneath the acquisition grid were similarly collected every $\mod(\Omega_N,5) =1$ and so on. Eventually, this permitted each projection to use 25 images with twenty-five different illuminations. Considering the detector field of view width of $2047$ pixels, $N = 1800$ projections were necessary to sample correctly an angular range of 180 degrees around the sample. After recovery of the phase images, a standard filtered-back projection was applied within the Radon transform inversion for CT.

Figure~\ref{fig:mureslice} shows slices of the blackberry reconstruction from the phase and absorption contrasts. The line cut in Fig.~\ref{fig:mureslice}.(d) shows the higher contrast rendered for the different grain densities parts obtained with phase contrast imaging as compared to the absorption modality.

\begin{figure}[t]
\ifpdf
\includegraphics{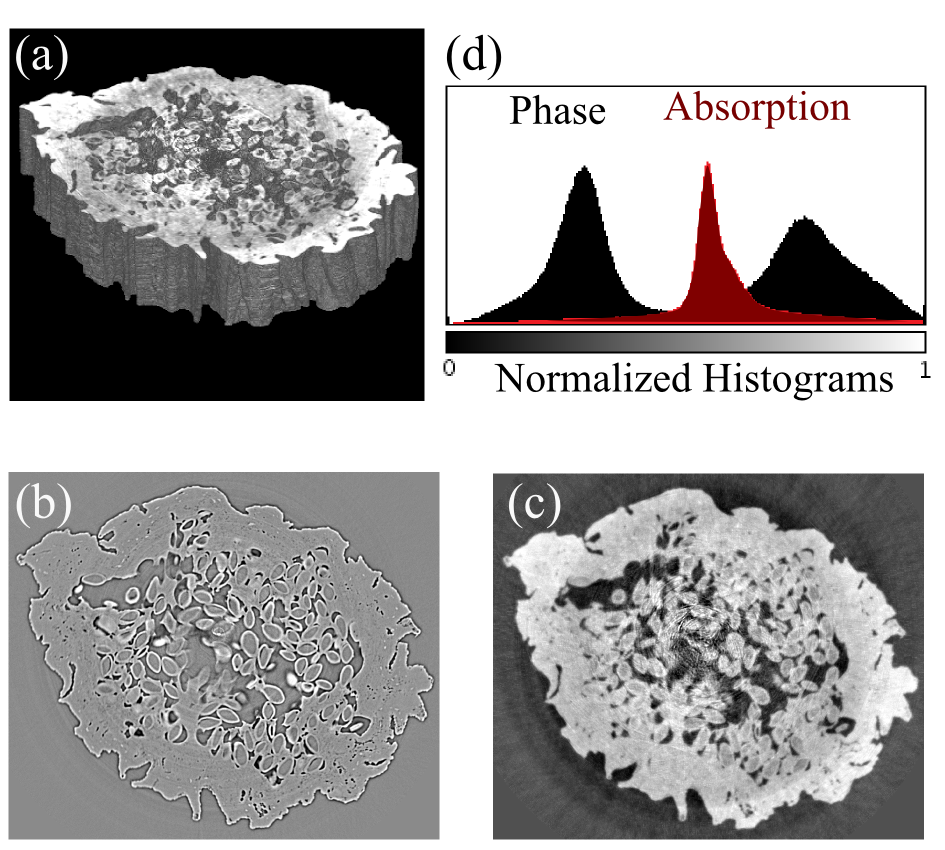}
\else
\includegraphics{vanilla.eps}
\fi
\caption{(Color online) (a) Volume rendering of a section of the dry vanilla bean. (b) Absorption contrast and (c) phase contrast slices reconstruction. (d) Normalized histogram of the slices shown in (b) and (c).\label{fig:boneslice}}
\end{figure}

The vanilla bean was also reconstructed from image scans collected using higher energy photons. Figure~\ref{fig:boneslice} shows volume and slice reconstructions of the sample. The histograms of the two images in Fig.~\ref{fig:boneslice}.(d), which were put to an equivalent amplitude and scale, show a larger distribution spread and separation of the voxel value for the phase image which eventually makes the segmentation easier.

These two sample reconstructions demonstrate the benefits of the method for CT. The moderate number of sample exposures makes the method already very interesting in regards to other available techniques. Besides, the reconstructions displayed here necessitated five images per projection while, in contrast, 40 images were used in the 1D XSS scheme of Ref.~\cite{wang2016apl}. In addition, the method presented here offers the advantage over 1DXSS of a better sensitivity for the second beam transverse direction and results in a more accurate phase recovery. This aspect turns out to be essential when one is operating a laboratory source with a moderate flux. Indeed the refraction sensitivity available with the XST, XSVT and 1D XSS techniques is here hampered by the use of photon counting detectors with a large pixel size.

\section{Conclusion}

We applied the 2D XSS method with a sparser sampling data collection scheme, thus reducing the overall number of images and making the process compatible with CT. The method was shown to work at both medium and high energy with large spectral bandwidth and provided high angular sensitivity and resolution.

The implementation of the method within a setup involving magnifying optics is expected to achieve the full potential of this full field imaging approach. Its application at laboratory low coherence sources will provide non-invasive ways to better depict the matter at the microscale. At synchrotrons it will eventually permit to achieve nanometer resolution phase contrast imaging as we intend to demonstrate it in future work. The next developments around the method will also aim at investigating ways of reducing the noise to ultimately optimize the number of sample exposures and define laws for the selection of the method parameters. As already performed with XSVT \cite{berujon2016prap} one could for instance envisage the consideration of information in neighboring pixel to reduce the number of images.

Finally, as the speckle pattern is nothing but a random modulator, it could be replaced by any phase or absorption pattern, such as a pseudo-random mask or coded apertures. In this context, such a processing method is applicable to other instruments based on wavefront modulation.

\onecolumngrid
\begin{acknowledgments}
The authors wish to thank P. Tafforeau for advice on the making of the filtered beam and the ESRF for financial and personal support.
\end{acknowledgments}
\twocolumngrid
%

\end{document}